\documentclass[twocolumn,showpacs,prb,citeautoscript,amsmath,amssymb,floatfix,10pt,aps]{revtex4-1}
\usepackage{amsmath,amssymb,color}
\usepackage{bm}
\usepackage{graphicx}
\usepackage{psfrag}
\newcommand{\bd}{\bm}

\begin{document}

\title{A rich man's derivation 
of scaling laws for the
 Kondo model}

\author{
Dmytro Tarasevych, Jan Krieg, and Peter Kopietz}
  
\affiliation{Institut f\"{u}r Theoretische Physik, Universit\"{a}t
  Frankfurt,  Max-von-Laue Strasse 1, 60438 Frankfurt, Germany}

\date{October 24, 2018}

 \begin{abstract}

We show how the one-loop ``poor man's 
scaling'' equations for the Kondo model with arbitrary impurity spin can be obtained within the 
framework of the
functional renormalization group approach for quantum spin systems
recently developed  by Krieg and 
Kopietz [arXiv:1807.02524]. 
We argue that our method supersedes the
``poor man's 
scaling'' approach and can also be used to study the strong coupling regime
of the Kondo model.

\end{abstract}

\maketitle

\section{Introduction}

In an influential paper entitled  {\it{``A poor man's derivation of scaling laws for the Kondo 
problem''}},  Anderson
has derived the scaling laws for the Kondo problem  using a cutoff renormalization 
technique.~\cite{Anderson70a}
Although these scaling laws have been derived
previously by means of a
complicated space-time approach~\cite{Anderson70b},  
Anderson's  approach
showed how these scaling laws emerge from the  renormalization group and
inspired many subsequent works.
In his  ``poor man's scaling'' approach,  Anderson
calculates the change of the 
$T$-matrix when  electronic
states with  momenta outside a shell around the Fermi surface are successively eliminated.
Given the fact that the $T$-matrix is an operator in Hilbert space,
the  operator renormalization  in the ``poor man's scaling'' approach
is conceptually different from the usual 
RG procedure  in classical statistical mechanics, where the  flow of coupling constants defined 
via a suitable  effective action is considered.
Although the ``poor man's scaling'' approach has been described
in the past 30 years in several  textbooks~[\onlinecite{Anderson84,Hewson93,Coleman15}], 
the technical details remain  somewhat cumbersome.

In this work we shall present an alternative method for deriving the
scaling equations for the Kondo model using the functional renormalization group (FRG) 
\cite{Wetterich93,Berges02,Kopietz10,Metzner12}.
We show that for a ``rich man'' equipped with this 
modern formulation of the Wilsonian renormalization 
group,  it is straightforward to obtain the 
scaling laws of the Kondo problem using a simple truncation of the average 
effective action.
We use a specific implementation of the FRG for quantum spin systems
recently developed in Ref.~[\onlinecite{Krieg18}], which does not require that the relevant degrees of freedom 
can be represented in terms of canonical fermions or bosons.
Apart from its formal elegance, this spin FRG approach has the advantage that 
it can also be used to study the strong coupling regime of the Kondo model where
Anderson's ``poor man's scaling'' approach is not valid.

\section{Exact renormalization group 
for the Kondo model}

The Kondo model describes a single localized spin $\bd{S}$, which
is coupled to a bath  of non-interacting conduction electrons 
in a metal \cite{Hewson93,Coleman15}.  The Kondo Hamiltonian is given by
\begin{align}
 {\cal{H}} &= \sum_{ij \sigma} t_{ij} c^{\dagger}_{i \sigma} c_{j \sigma} + J \bd{S} 
 \cdot \bd{s}_0
 \nonumber
 \\
&=
 \sum_{\bd{k} \sigma} \epsilon_{\bd{k}} c^{\dagger}_{\bd{k} \sigma}
 c_{\bd{k} \sigma} + J \bd{S} \cdot  \bd{s}_0,
 \label{eq:kondo2}
 \end{align}
where the
operators $c^{\dagger}_{i \sigma}$ and $c_{i \sigma}$ create and
annihilate an electron with spin $\sigma = \uparrow , \downarrow$ at lattice site $i$,
while the operator $\bd{S}$ represents an additional spin with magnitude $S$ which
couples via the exchange coupling $J$ to the electronic spin $\bd{s}_0$ at the origin.
In the second line of Eq.~(\ref{eq:kondo2}) we have transformed the electronic part 
of the Hamiltonian to momentum space,
 \begin{equation}
 c_{i \sigma} = \frac{1}{\sqrt{N}} \sum_{\bd{k}} e^{ i \bd{k} \cdot \bd{r}_i } c_{\bd{k} \sigma},
 \end{equation}
where $N$ is the number of lattice sites and the energy dispersion $\epsilon_{\bd{k}}$ is related to the hopping 
energies $t_{ij}$ via
 \begin{equation}
 \epsilon_{\bd{k}} = \frac{1}{N} \sum_{ij} e^{ - i  \bd{k} \cdot ( \bd{r}_i - \bd{r}_j ) } t_{ij}.
 \end{equation}
The electronic spin at the site of the impurity spin is explicitly given by
 \begin{equation}
 \bd{s}_0 = \sum_{\sigma \sigma^{\prime}} c^{\dagger}_{i=0, \sigma} 
 \bd{\sigma}_{ \sigma \sigma^{\prime}} c_{ i=0, \sigma^{\prime}} =
 \frac{1}{N}  \sum_{ \bd{k} \bd{k}^{\prime}} \sum_{\sigma \sigma^{\prime}}
 c^{\dagger}_{\bd{k} \sigma} {\bd{\sigma}}_{\sigma \sigma^{\prime}} 
 c_{\bd{k}^{\prime} \sigma^{\prime}},
 \end{equation}
where $\bd{\sigma}$ is the vector of Pauli matrices.
Note that with our normalization, the exchange coupling $J$ has units of energy and the density of states at the Fermi energy $\epsilon_F$,
\begin{equation}
 \rho_0 = \frac{1}{N} \sum_{\bd{k}} \delta ( \epsilon_{\bd{k}} - \epsilon_F ),
 \label{eq:rho0def}
 \end{equation}
has units of inverse energy.
The components of the spin operator $\bd{S}$ satisfy the
$SU(2)$ commutation relations
 \begin{equation}
 [ S^{\alpha } , S^{\beta} ] = i \epsilon_{\alpha \beta \gamma} S^{\gamma},
 \end{equation}
where the superscripts $\alpha, \beta, \gamma$ refer to the Cartesian components of the vector operator $\bd{S}$
and $\epsilon_{\alpha \beta \gamma}$ is the
totally antisymmetric $\epsilon$-tensor. 
Although in his ``poor man's scaling'' approach
Anderson studied only the case $S=1/2$, within our spin FRG 
it is straightforward to consider an arbitrary impurity spin $S$. 
Since the spin commutation relations are neither bosonic nor fermionic,
the usual machinery of
many-body perturbation theory 
is not directly applicable to the Kondo model.
A popular solution to this problem is to represent the 
spin operators in terms of Abrikosov 
pseudofermions \cite{Abrikosov65}, 
but this requires an additional projection to eliminate the
unphysical part of the enlarged Hilbert space \cite{Coleman15}. 
The crucial insight of our recent work \cite{Krieg18} is that the
powerful FRG formalism can be directly applied to quantum spin systems
without using any representation of the spin operators in terms
of auxiliary variables. One simply has to define a proper cutoff scheme and
write down the relevant generating functional in terms of a trace over the 
physical Hilbert space. For classical spin models and bosonic quantum lattice models,
a similar strategy has been adopted earlier in Refs.~[\onlinecite{Machado10,Rancon11A,*Rancon11B,*Rancon12A,*Rancon12B,Rancon14}].

To implement the bandwidth cutoff used in Anderson's ``poor man's scaling'' approach,
we consider  the following cutoff-dependent deformation of the Kondo 
Hamiltonian, 
 \begin{equation}
 {\cal{H}}_{\Lambda} = {\cal{H}}_0 + {\cal{V}}_{\Lambda},
 \end{equation}
where ${\cal{H}}_0$ represents the local exchange interaction,
  \begin{equation}
 {\cal{H}}_0 = J {\bd{S}}  \cdot \bd{s}_0,
 \end{equation}
while the cutoff-dependent operator  ${\cal{V}}_{\Lambda}$ represents the 
regularized electronic kinetic energy,
\begin{equation}
 {\cal{V}}_{\Lambda} =  \sum_{\bd{k} \sigma} [ \xi_{\bd{k}} + R_{\Lambda} ( \bd{k} ) ] 
c^{\dagger}_{\bd{k} \sigma}
 c_{\bd{k} \sigma},
 \label{eq:electronscutoff}
 \end{equation}
where the dispersion $\xi_{\bd{k}} = \epsilon_{\bd{k}} - \epsilon_F$ is measured relative to the Fermi energy $\epsilon_F$.
The regulator function $R_{\Lambda} ( \bd{k} ) $ should be chosen such that 
it vanishes for $\Lambda \rightarrow 0$, so that in this limit we recover 
our original model. On the other hand, at some large  initial cutoff scale
$\Lambda_0$  the contribution from states with energies in some interval around the Fermi energy
should be suppressed. 
A possible regulator  with this property is \cite{Litim01,Machado10,Krieg17}
 \begin{equation}
 R_{\Lambda} ( \bd{k} ) =   {\rm{sign}} ( \xi_{\bd{k}} )  ( \Lambda -   | \xi_{\bd{k}} | )
\Theta ( \Lambda -   | \xi_{\bd{k}} | ).
 \label{eq:Litim}
 \end{equation}
The initial value $\Lambda_0$ of the cutoff should be identified with the total
bandwidth of the dispersion, $\Lambda_0 = {\rm max}_{\bd{k}} \{ | \xi_{\bd{k}} | \}.$

To derive exact FRG flow equations for the Kondo model, let us consider the
generating functional of the cutoff-dependent connected correlation functions,
 \begin{eqnarray}
& &  e^{ {\cal{G}}_{\Lambda} [  \bar{\eta}, \eta , \bd{h}  ] } 
 \nonumber
 \\
& = & {\rm Tr} \left[
 e^{ - \beta {\cal{H}}_0 } {\cal{T}} 
 e^{  ( \bar{\eta},  c ) + ( c^{\dagger} , \eta ) + \int_0^{\beta} d \tau 
 [ \bd{h} ( \tau )  \cdot {\bd{S}} ( \tau )  -  {\cal{V}}_{\Lambda} ( \tau)]}
 \right],
 \label{eq:GLdef}
 \hspace{7mm}
 \end{eqnarray}
where the trace is over the electronic Fock space as well as over the Hilbert space of the localized spin
and
we have used the following short notation for the source terms 
of the electron operators, 
 \begin{eqnarray}
 & & ( \bar{\eta} , c ) + ( c^{\dagger} , \eta )  
 \nonumber
 \\
 & =  &
\int_0^{\beta} d \tau 
 \sum_{\bd{k} \sigma } \left[  \bar{\eta}_{\bd{k} \sigma} ( \tau )  c_{\bd{k} \sigma }  ( \tau )
+ c^{\dagger}_{\bd{k} \sigma } ( \tau )   \eta_{\bd{k} \sigma} (\tau ) \right].
 \end{eqnarray}
Here the sources $\bar{\eta}_{\bd{k} \sigma} ( \tau )$ and
$\eta_{\bd{k} \sigma} ( \tau )$ are Grassmann variables,
$\bd{h} ( \tau )$ is a time-dependent external magnetic field,
$\beta$ is the inverse temperature and 
${\cal{T}}$ is the Wick time-ordering operator in imaginary time, 
i.e., operators at larger values of $\tau$ in the expansion of the exponential should be moved to the left, with extra minus signs generated by the 
permutation of any pair of fermion operators.
The time evolution of all operators under the time-ordering symbol
in Eq.~(\ref{eq:GLdef}) is in the
imaginary-time
interaction picture with respect to ${\cal{H}}_0$, for example
 \begin{equation}
 {\bd{S}} ( \tau ) = e^{ {\cal{H}}_0 \tau } {\bd{S}}  e^{ - {\cal{H}}_0 \tau }.
 \end{equation}
Taking variational derivatives of ${\cal{G}}_{\Lambda} [ \bar{\eta} , \eta, \bd{h} ]$ 
with respect to the sources  and then setting the sources equal to zero, we can obtain 
all types of
time-ordered connected 
correlation functions of the Kondo model.
In particular, the magnetic moment of the localized  spin is
 \begin{eqnarray}
  \bd{m}_{\Lambda} & = &
 \left.
\frac{ \delta {\cal{G}}_{\Lambda} [ 0,0 , \bd{h} ] }{\delta \bd{h} ( \tau ) }
 \right|_{\bd{h} =0} 
  =  \langle  \bd{S}  \rangle ,
 \label{eq:mdef}
\end{eqnarray}
and the time-ordered two-point connected correlation function of the localized spin is
 \begin{eqnarray}
 F^{\alpha \beta}_{\Lambda} ( \tau - \tau^{\prime} ) & = &
 \left.
 \frac{ \delta^2 {\cal{G}}_{\Lambda} [ 0 , 0 ,  \bd{h} ] }{\delta h_{\alpha} ( \tau )   \delta h_{\beta} ( \tau^{\prime} ) } 
 \right|_{\bd{h} =0}
 \nonumber
 \\
 &=& \langle {\cal{T}} \bigl( S^{\alpha} ( \tau ) 
 S^{\beta} ( \tau^{\prime} ) \bigr)
\rangle   -  \langle S^{\alpha}  \rangle
 \langle  S^{\beta}  \rangle,
 \label{eq:Gijdef}
 \hspace{7mm}
 \end{eqnarray}
where $ \langle \ldots \rangle$ denotes the equilibrium expectation value
with the deformed Hamiltonian ${\cal{H}}_{\Lambda}$ for vanishing sources, 
 \begin{eqnarray}
 & & \langle \ldots  \rangle =
 \frac{  {\rm Tr} \left[ e^{ - \beta {\cal{H}}_{\Lambda}    }  \ldots \right]  }{
 {\rm Tr} \left[ e^{ - \beta  {\cal{H}}_{\Lambda} }  \right] }.
 \label{eq:averagedef}
 \hspace{7mm}
 \end{eqnarray}
Similarly, by taking  derivatives with respect to the Grassmann sources $\eta$ and $\bar{\eta}$, 
we obtain the time-ordered connected correlation functions of the conduction electrons.
For example, the electronic single-particle Green function is given by
 \begin{eqnarray}
 G^{\sigma}_{\Lambda} ( \bd{k} , {\bd{k}}^{\prime},  
\tau - \tau^\prime ) & = & \left. \frac{ \delta^2 {\cal{G}}_{\Lambda} [ \bar{\eta},
 \eta , 0 ]}{\delta \bar{\eta}_{\bd{k} \sigma} ( \tau )
 \delta \eta_{ \bd{k}^{\prime} \sigma} ( \tau^{\prime} ) } 
 \right|_{ \eta = \bar{\eta} =0} 
 \nonumber
 \\
 & = &  - \langle {\cal{T}} [ c_{\bd{k} \sigma} ( \tau )
 c^{\dagger}_{\bd{k}^{\prime} \sigma} ( \tau^{\prime} ) ] \rangle.
 \end{eqnarray}

Following Ref.~[\onlinecite{Krieg18}], we can
derive an exact flow equation for the generating functional
 ${\cal{G}}_{\Lambda} [  \bar{\eta}, \eta , \bd{h}  ]$ by simply
taking the derivative of Eq.~(\ref{eq:GLdef})  
with respect to the cutoff parameter $\Lambda$. We then obtain the
exact FRG flow equation
 \begin{eqnarray}
 & & \partial_{\Lambda}{\cal{G}}_{\Lambda} [ \bar{\eta} , \eta , \bd{h} ]  
=  \int_0^{\beta} d \tau \sum_{\bd{k} \sigma}  ( \partial_{\Lambda} R_{\Lambda} ( \bd{k} ) )
 \biggl[
 \frac{ \delta^2 {\cal{G}}_{\Lambda} [ \bar{\eta} , \eta , \bd{h} ]  }{ \delta \eta_{\bd{k} \sigma}
 ( \tau ) \delta \bar{\eta}_{\bd{k} \sigma} ( \tau ) }
 \nonumber
 \\
 & & \hspace{35mm} +
     \frac{ \delta {\cal{G}}_{\Lambda} [ \bar{\eta} , \eta , \bd{h} ]  }{ 
 \delta \eta_{\bd{k} \sigma}     ( \tau ) }
     \frac{ \delta {\cal{G}}_{\Lambda} [ \bar{\eta} , \eta , \bd{h} ]  }{
 \delta \bar{\eta}_{\bd{k} \sigma} ( \tau ) } \biggr].
 \label{eq:flowG}
 \end{eqnarray}
Finally, we introduce the generating functional of the cutoff-dependent irreducible vertices,
 \begin{eqnarray}
 & & \Gamma_{\Lambda} [ \bar{\psi} , \psi , \bd{m} ]  =
 ( \bar{\eta} , \psi ) + ( \bar{\psi} , \eta ) +
\int_0^{\beta} d \tau \bd{h} ( \tau ) \cdot \bd{m} ( \tau )
 \nonumber
 \\
 & & 
  - {\cal{G}}_{\Lambda} [ \bar{\eta} , \eta , \bd{h}  ]
 -   \int_0^{\beta} d \tau \sum_{ \bd{k} \sigma } R_{\Lambda} ( \bd{k} ) 
  \bar{\psi}_{ \bd{k} \sigma} ( \tau )
  \psi_{\bd{k} \sigma} ( \tau ),
 \label{eq:GammaLambdadef}
  \end{eqnarray}
which differs from the usual Legendre transform by the subtraction of the
regulator term. On the right-hand side of Eq.~(\ref{eq:GammaLambdadef}) it is understood
that the sources 
${\eta}_{\bd{k} \sigma } ( \tau ) = {\eta}_{\bd{k} \sigma } ( \tau ; \bar{\psi} , \psi , \bd{m} )$,
$\bar{\eta}_{\bd{k} \sigma } ( \tau ) = \bar{\eta}_{\bd{k} \sigma } ( \tau ; \bar{\psi} , \psi , \bd{m} )$, and
$\bd{h} ( \tau ) = \bd{h} ( \tau ; \bar{\psi} , \psi , \bd{m} )$
should be considered as functionals of the expectation values 
$\psi_{\bd{k} \sigma} ( \tau ) = \langle {\cal{T}} c_{\bd{k} \sigma} ( \tau ) \rangle$,
$\bar{\psi}_{\bd{k} \sigma} ( \tau ) = \langle {\cal{T}} c^{\dagger}_{\bd{k} \sigma} ( \tau ) \rangle$, and
$\bd{m} ( \tau ) = \langle {\cal{T}} \bd{S} ( \tau ) \rangle$ 
by inverting the relations 
 \begin{eqnarray}
 \psi_{\bd{k} \sigma} ( \tau ; \bar{\eta} , \eta, \bd{h}  ) & = & 
  \langle {\cal{T} } c_{\bd{k} \sigma} ( \tau ) \rangle = \frac{  \delta {\cal{G}}_{\Lambda} [ \bar{\eta} , \eta , \bd{h} ] }{
 \delta \bar{\eta}_{\bd{k} \sigma} ( \tau ) },
 \\
 \bar{\psi}_{\bd{k} \sigma} ( \tau ; \bar{\eta} , \eta, \bd{h}  ) & = & 
  \langle {\cal{T}} c^{\dagger}_{\bd{k} \sigma} ( \tau ) \rangle = - \frac{  \delta {\cal{G}}_{\Lambda} [ \bar{\eta} , \eta , \bd{h} ] }{
 \delta {\eta}_{\bd{k} \sigma} ( \tau ) },
 \\
  m^{\alpha} ( \tau ; \bar{\eta} , \eta , \bd{h} ) & = &
  \langle {\cal{T}} S^{\alpha} ( \tau ) \rangle 
 = \frac{ \delta {\cal{G}}_{\Lambda} [ \bar{\eta}, \eta , \bd{h} ] }{\delta h_{\alpha} ( \tau ) } .
\end{eqnarray}
Note that by construction, the sources as functionals of the expectation values
can be obtained from the derivatives of
the functional  $\Gamma_{\Lambda} [ \bar{\psi}, \psi, \bd{m} ]$ as follows,
 \begin{eqnarray}
 \eta_{\bd{k} \sigma} ( \tau ; \bar{\psi} , \psi, \bd{m}  ) - R_{\Lambda} ( \bd{k} )  
 \psi_{\bd{k} \sigma} ( \tau )
 & = & 
  \frac{  \delta \Gamma_{\Lambda} [ \bar{\psi} , \psi , \bd{m} ] }{
 \delta \bar{\psi}_{\bd{k} \sigma} ( \tau ) },
 \\
 \bar{\eta}_{\bd{k} \sigma} ( \tau ; \bar{\psi} , \psi, \bd{m}  ) 
- R_{\Lambda} ( \bd{k} )  
 \bar{\psi}_{\bd{k} \sigma} ( \tau )
& = & 
  - \frac{  \delta \Gamma_{\Lambda}  [ \bar{\psi} , \psi , \bd{m} ] }{
 \delta {\psi}_{\bd{k} \sigma} ( \tau ) },
 \hspace{7mm}
\\
h_{\alpha} ( \tau ; \bar{\psi} , \psi , \bd{m} ) & = &
 \frac{ \delta \Gamma_{\Lambda} [ \bar{\psi}, \psi, \bd{m} ] }{\delta m^{\alpha} ( \tau ) }.
 \end{eqnarray}
Taking the derivative of Eq.~(\ref{eq:GammaLambdadef}) with respect to $\Lambda$ and using the flow equation~(\ref{eq:flowG}), we find that
$ \Gamma_{\Lambda} [ \bar{\psi} , \psi , \bd{m} ]$ satisfies
 \begin{eqnarray}
\partial_{\Lambda} \Gamma_{\Lambda} [ \bar{\psi} , \psi , \bd{m} ] & = &
 -  \int_0^{\beta} d \tau \sum_{\bd{k} \sigma}  ( \partial_{\Lambda} R_{\Lambda} ( \bd{k} ) )
 \nonumber
 \\
 & & \times
\frac{ \delta^2 {\cal{G}}_{\Lambda} [ \bar{\eta} , \eta , \bd{h} ]  }{ \delta \eta_{\bd{k} \sigma}
 ( \tau ) \delta \bar{\eta}_{\bd{k} \sigma} ( \tau ) }.
 \label{eq:flowGamma1}
 \end{eqnarray}
At this point it is convenient to collect all fields into a seven-component superfield
$(\Phi_{\alpha})  = ( \psi_{\uparrow} , \bar{\psi}_{\uparrow} ,     \psi_{\downarrow} , \bar{\psi}_{\downarrow}  , \bd{m} )$, where the label $\alpha$ denotes all parameters which are necessary to specify the field configuration, including the field type.
The functional derivative in the last line of Eq.~(\ref{eq:flowGamma1})
can then be transformed  as follows,
 \begin{eqnarray}
& & \frac{ \delta^2 {\cal{G}}_{\Lambda} [ \bar{\eta} , \eta , \bd{h} ]  }{ \delta \eta_{\bd{k} \sigma}
 ( \tau ) \delta \bar{\eta}_{\bd{k} \sigma} ( \tau ) }
 \nonumber
 \\
 &=  &   \left[ \frac{ \delta}{ \delta \Phi } \otimes \frac{ \delta }{\delta \Phi }  
{{\Gamma}}_{\Lambda} [ \Phi ]     + \mathbf{Z} \mathbf{R}_{\Lambda} \right]^{-1}_{ \alpha = (\bar{\psi} , \bd{k} , \sigma, \tau ), 
 \alpha^{\prime} = ( \psi, \bd{k}, \sigma , \tau ) }
 \nonumber
 \\
 & = & 
\left[   \mathbf{Z}\left( \mathbf{Z} \frac{ \delta}{ \delta \Phi } \otimes \frac{ \delta }{\delta \Phi }  
{{\Gamma}}_{\Lambda} [ \Phi ]     +  \mathbf{R}_{\Lambda} \right) \right]^{-1}_{ \alpha = (\bar{\psi} , \bd{k} , \sigma, \tau ), 
 \alpha^{\prime} = ( \psi, \bd{k}, \sigma , \tau ) }
\nonumber
 \\
 & = & 
\left[   \mathbf{Z}\left( \left[ \frac{ \delta}{ \delta \Phi } \otimes 
 \frac{ \delta }{\delta \Phi }   \right]^T
{{\Gamma}}_{\Lambda} [ \Phi ]    +  \mathbf{R}_{\Lambda} \right) \right]^{-1}_{ \alpha = (\bar{\psi} , \bd{k} , \sigma, \tau ), 
 \alpha^{\prime} = ( \psi, \bd{k}, \sigma , \tau ) }
\nonumber
 \\
 & = & \left[   \mathbf{Z}\left( 
{ \mathbf{\Gamma}}^{\prime \prime}_{\Lambda} [ \Phi ]     +  \mathbf{R}_{\Lambda} \right) \right]^{-1}_{ \alpha = (\bar{\psi} , \bd{k} , \sigma, \tau ), 
 \alpha^{\prime} = ( \psi, \bd{k}, \sigma , \tau ) }
 \nonumber
 \\
 & = & \left[  \left( 
 { \mathbf{\Gamma}}^{\prime \prime}_{\Lambda} [ \Phi ]     +  \mathbf{R}_{\Lambda} 
 \right)^{-1}   \mathbf{Z}  \right]_{ \alpha = (\bar{\psi} , \bd{k} , \sigma, \tau ), 
 \alpha^{\prime} = ( \psi, \bd{k}, \sigma , \tau ) }
\nonumber
 \\
 & = & 
\left[  
 { \mathbf{\Gamma}}^{\prime \prime}_{\Lambda} [ \Phi ]     +  \mathbf{R}_{\Lambda} 
  \right]^{-1}_{ \alpha = ({\psi} , \bd{k} , \sigma, \tau ), 
 \alpha^{\prime} = ( \bar{\psi}, \bd{k}, \sigma , \tau ) },
 \end{eqnarray}
where the matrix elements of the derivative operator $\frac{ \delta}{ \delta \Phi } \otimes \frac{ \delta }{\delta \Phi }  $ are defined by
 \begin{equation}
 \left[ \frac{ \delta}{ \delta \Phi } \otimes \frac{ \delta }{\delta \Phi }  
 \right]_{\alpha \beta} = \frac{ \delta }{ \delta \Phi_{\alpha} }  
 \frac{ \delta }{ \delta \Phi_{\beta} }, \; \; \; 
 \left[ \frac{ \delta}{ \delta \Phi } \otimes \frac{ \delta }{\delta \Phi }  
 \right]^T_{\alpha \beta} = \frac{ \delta }{ \delta \Phi_{\beta} }  
 \frac{ \delta }{ \delta \Phi_{\alpha} }.
 \end{equation} 
The statistics matrix $\mathbf{Z}$ is diagonal, $\mathbf{Z}_{\alpha \alpha^{\prime}}
 = \delta_{\alpha \alpha^{\prime} } \zeta_{\alpha}$, where $\zeta_{\alpha} =1 $ if $\alpha$  refers to one of the components of the
magnetization field $\bd{m}$
 and $\zeta_{\alpha} =-1$ if $\alpha$ labels a fermionic field.
The regulator matrix $\mathbf{R}_{\Lambda}$ is defined by writing
the regulator term in superfield notation as
 \begin{eqnarray}
 & & 
\int_0^{\beta} d \tau \sum_{ \bd{k} \sigma } R_{\Lambda} ( \bd{k} ) 
  \bar{\psi}_{ \bd{k} \sigma} ( \tau )
  \psi_{\bd{k} \sigma} ( \tau )
= \frac{1}{2} \int_{\alpha} \int_{\beta} \Phi_{\alpha} [ \mathbf{R}_{\Lambda} ]_{\alpha \beta } \Phi_{\beta},
 \nonumber
 \\
 & &
 \end{eqnarray}
and the matrix $ { \mathbf{\Gamma}}^{\prime \prime}_{\Lambda} [ \Phi ] $
is defined by
 \begin{eqnarray}
 { \mathbf{\Gamma}}^{\prime \prime}_{\Lambda} [ \Phi ] 
&  = &
  \mathbf{Z} \frac{ \delta}{ \delta \Phi } \otimes \frac{ \delta }{\delta \Phi }  
{{\Gamma}}_{\Lambda} [ \Phi ]  =
\left(  \frac{ \delta}{ \delta \Phi } \otimes \frac{ \delta }{\delta \Phi }  
{{\Gamma}}_{\Lambda} [ \Phi ] \right)^T
\nonumber
 \\
 & = &
\left(  \frac{ \delta}{ \delta \Phi } \otimes \frac{ \delta }{\delta \Phi }   \right)^T
{{\Gamma}}_{\Lambda} [ \Phi ].
 \end{eqnarray}
The flow equation  (\ref{eq:flowGamma1}) can therefore be written in the  
compact matrix form 
 \begin{eqnarray}
 \partial_{\Lambda} \Gamma_{\Lambda} [ \Phi ]  &=  &
\frac{1}{2} {\rm STr} \left\{   ( \partial_{\Lambda} \mathbf{R}_{\Lambda} )
 \left[  \mathbf{\Gamma}^{\prime \prime}_{\Lambda} [ \Phi ]   + \mathbf{R}_{\Lambda} \right]^{-1} \right\} 
 \nonumber
 \\
 &  & \hspace{-20mm} =
 \frac{1}{2} {\rm STr} \left\{   ( \partial_{\Lambda} \mathbf{R}_{\Lambda} )
 \left[  \left( \frac{ \delta}{ \delta \Phi } \otimes \frac{ \delta }{\delta \Phi } 
  \right)^T   \Gamma_{\Lambda} [ \Phi ]   + \mathbf{R}_{\Lambda} \right]^{-1} \right\},
 \nonumber
 \\
 & &
 \label {eq:WetterichKondo}
 \end{eqnarray}
where the supertrace is defined by ${\rm STr} \{ \ldots \} = {\rm Tr} \{ \mathbf{Z} \ldots \}$. Eq.~(\ref{eq:WetterichKondo}) is a generalized  Wetterich equation \cite{Wetterich93} for the Kondo model. As already emphasized in Ref.~[\onlinecite{Krieg18}],
the spin degrees of freedom appear in Eq.~(\ref{eq:WetterichKondo}) in the same way as a bosonic field, which is a consequence of the fact that time-ordered spin correlation functions satisfy bosonic Kubo-Martin-Schwinger boundary conditions.
We emphasize that the FRG flow equation (\ref{eq:WetterichKondo}) is formally exact
and encodes the renormalization group flow of all correlation functions of the
Kondo model.

\section{Scaling equations for the anisotropic spin-$S$ Kondo model}

\begin{figure}
\vspace{7mm}
\centering\includegraphics[width=0.35\textwidth]{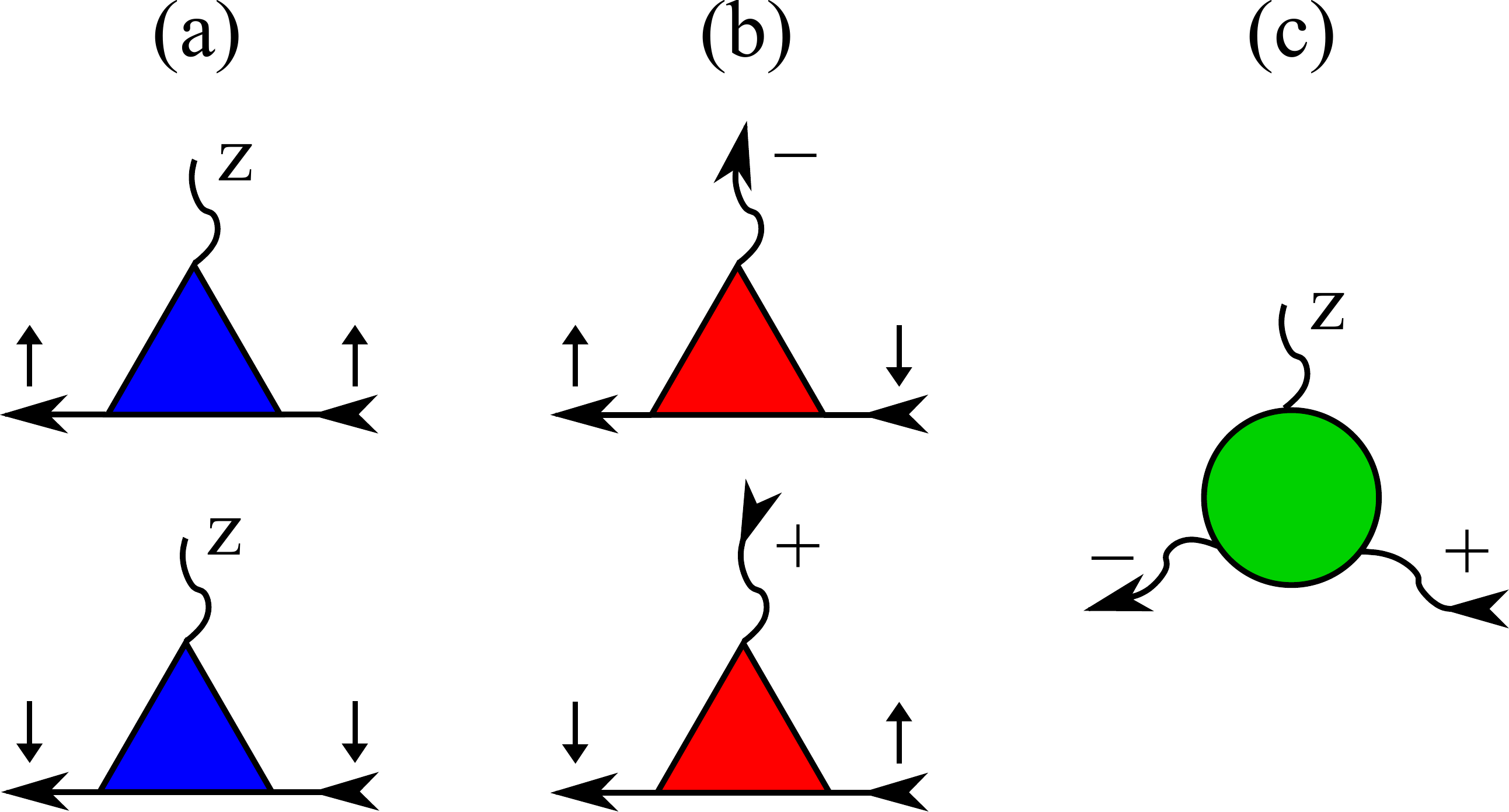}
\caption{
Graphical representation of the
different types of vertices with three external  legs, which appear
in the expansion of the average effective action for the anisotropic  Kondo model
given in Eq.~(\ref{eq:Gammaaniso2}).
(a) Longitudinal part of the exchange interaction, $J^{\uparrow \uparrow }_{\Lambda}
 = J^{\parallel}_{\Lambda}$ and $J^{\downarrow \downarrow }_{\Lambda}
 = - J^{\parallel}_{\Lambda}$.
 (b) Transverse part of the 
exchange interaction, $J^{\uparrow \downarrow}_{\Lambda} = J^{\downarrow \uparrow}_{\Lambda} = J^{\bot}_{\Lambda}$.
(c) Interaction vertex $\Gamma^{-+z}_{\Lambda}$
describing dynamic correlations of the
impurity spin. Solid outgoing arrows represent $\bar{\psi}_{\bd{k} \sigma}$ and solid incoming arrows represent $\psi_{ \bd{k} \sigma}$, with spin projections 
given by the arrows above the lines. Wavy lines without an arrow represent $m^z$,
wavy lines with an outgoing arrow represent $m^-$, and wavy lines with an incoming arrow represent $m^+$.
}
\label{fig:vertices}
\end{figure}

In this section we show how to obtain from Eq.~(\ref{eq:WetterichKondo})
the leading-order
scaling equations for the Kondo model for arbitrary impurity spin $S$.
It is instructive to consider  here the more general anisotropic Kondo
Hamiltonian,
 \begin{equation}
 {\cal{H}} = \sum_{\bd{k} \sigma} \epsilon_{\bd{k}} c^{\dagger}_{\bd{k} \sigma}
 c_{\bd{k} \sigma} + J^{\parallel} {S}^z {s}^z_0 + J^{\bot} ( S^x s^x_0 + S^y s^y_0 ).
 \label{eq:kondo3}
 \end{equation}
By expanding the generating functional
 $\Gamma_{\Lambda} [ \Phi ] = \Gamma_{\Lambda} [ \bar{\psi} , \psi , \bd{m} ]$ in powers of the fields, we can reduce the flow equation
(\ref{eq:WetterichKondo}) to an infinite hierarchy of coupled flow equations for the
irreducible vertices. For the anisotropic Kondo model, the expansion of 
 $\Gamma_{\Lambda} [ \bar{\psi} , \psi , \bd{m} ]$ up to third order in the fields is of the form
\begin{widetext}
 \begin{eqnarray}
 \Gamma_{\Lambda} [ \bar{\psi} , \psi, \bd{m}  ]   & = &  \Gamma^{(0)}_{\Lambda}
 +  \frac{1}{ N \beta} \sum_{  \bd{k}  \bd{k}^{\prime}  \sigma} \sum_{\omega}
 \left[ 
     N \delta_{ \bd{k} , \bd{k}^{\prime}}   
( \xi_{\bd{k}  } - i \omega )     +    \Sigma_{\Lambda} (   \bd{k} , \bd{k}^{\prime} ,  \omega )  \right]
  \bar{\psi}_{ \bd{k} \omega \sigma }  \psi_{
 \bd{k}^{\prime} \omega  \sigma }
 \nonumber
 \\
 & + &  \frac{1}{\beta} \sum_{ \nu} \left[   \frac{1}{2} [ F^{\parallel}_{\Lambda} (   \nu ) ]^{-1}  
{m}^z_{- \nu} {m}^z_{\nu } + [ F^{\bot}_{\Lambda} (  \nu ) ]^{-1}
 m^-_{  -\nu } m^+_{ \nu} \right]
 \nonumber
 \\
 &   + &  \frac{1}{ N \beta^2} \sum_{ \bd{k} \bd{k}^{\prime}  } \sum_{ \omega \omega^{\prime}} 
  J^{\parallel}_{\Lambda} ( \bd{k} ,  \omega , \bd{k}^{\prime} ,   \omega^{\prime} )
 [  \bar{\psi}_{ \bd{k} \omega \uparrow} 
 \psi_{ \bd{k}^{\prime}  \omega^{\prime} \uparrow} 
 -  \bar{\psi}_{ \bd{k} \omega \downarrow} 
 \psi_{ \bd{k}^{\prime}  \omega^{\prime} \downarrow}  ]
{m}^z_{ \omega - \omega^{\prime}} 
 \nonumber
 \\
& +  &  \frac{1}{ N \beta^2} \sum_{ \bd{k} \bd{k}^{\prime}}  \sum_{ \omega \omega^{\prime}} 
 J_{\Lambda}^{\bot} ( \bd{k} ,  \omega , \bd{k}^{\prime} ,   \omega^{\prime} )
 \bigl[ \bar{\psi}_{ \bd{k} \omega \uparrow} 
 \psi_{ \bd{k}^{\prime}  \omega^{\prime} \downarrow} 
  {m}^-_{ \omega - \omega^{\prime}} 
 + \bar{\psi}_{ \bd{k} \omega \downarrow} 
 \psi_{ \bd{k}^{\prime}  \omega^{\prime} \uparrow} 
  {m}^+_{ \omega - \omega^{\prime}} \bigr]
 \nonumber
 \\
& + & \frac{1}{\beta^2} \sum_{ \nu_1  \nu_2  \nu_3  }  \delta_{ \nu_1 + \nu_2 + \nu_3  ,0}
 \Gamma^{-+z}_{\Lambda} ( \nu_1 , \nu_2 , \nu_3   ) m^-_{\nu_1} m^+_{\nu_2 }
 m^z_{\nu_3 } + \ldots,
 \label{eq:Gammaaniso2}
 \end{eqnarray}
\end{widetext}
where the ellipsis represents terms involving more than three powers of the fields
and the transverse magnetization is expressed in terms of its spherical components
$m^{\pm}_{\nu} = m^x_{\nu} \pm i m^y_{\nu}$.
The Fourier transform to frequency space is defined as follows,
 \begin{eqnarray}
 \psi_{\bd{k} \sigma} ( \tau ) & =  &  \frac{1}{\beta} \sum_{\omega} e^{ - i \omega \tau } \psi_{ 
 \bd{k} \omega \sigma },
 \\
  \bd{m}  ( \tau ) & =  &  \frac{1}{\beta} \sum_{\nu} e^{ - i \nu \tau } \bd{m}_{ \nu },
 \end{eqnarray}
where $ \omega = 2 \pi ( n + 1/2) T$ is a fermionic Matsubara frequency and
$\nu = 2 \pi n T$ is a bosonic one.
In the first line of Eq.~(\ref{eq:Gammaaniso2}), the function
$\Sigma_{\Lambda} ( \bd{k} , \bd{k}^{\prime} ,  \omega )$
is the electronic self-energy generated by the coupling to the impurity spin.
In the second line, the functions $F^{\parallel}_{\Lambda} ( \nu )$ and
$F^{\bot}_{\Lambda} ( \nu )$ can be identified with the longitudinal and the transverse 
dynamic susceptibility of the impurity spin.
The functions $J^{\parallel}_{\Lambda} ( \bd{k} , \omega , \bd{k}^{\prime} ,
 \omega^{\prime} ) $ and
$J^{\bot}_{\Lambda} ( \bd{k} , \omega , \bd{k}^{\prime} ,
 \omega^{\prime} ) $ can be identified with renormalized longitudinal and transverse exchange interactions, which in general depend on the momenta and frequencies of the incoming and outgoing fermions. Finally, the vertex
$\Gamma^{-+z}_{\Lambda} ( \nu_1 , \nu_2 , \nu_3   )$ describes dynamic correlations between the components of the impurity spin.
We represent the  different types of three-legged vertices in the expansion~(\ref{eq:Gammaaniso2}) by the graphical symbols shown  in Fig.~\ref{fig:vertices}.

\begin{figure*}
\centering\includegraphics[width=\textwidth]{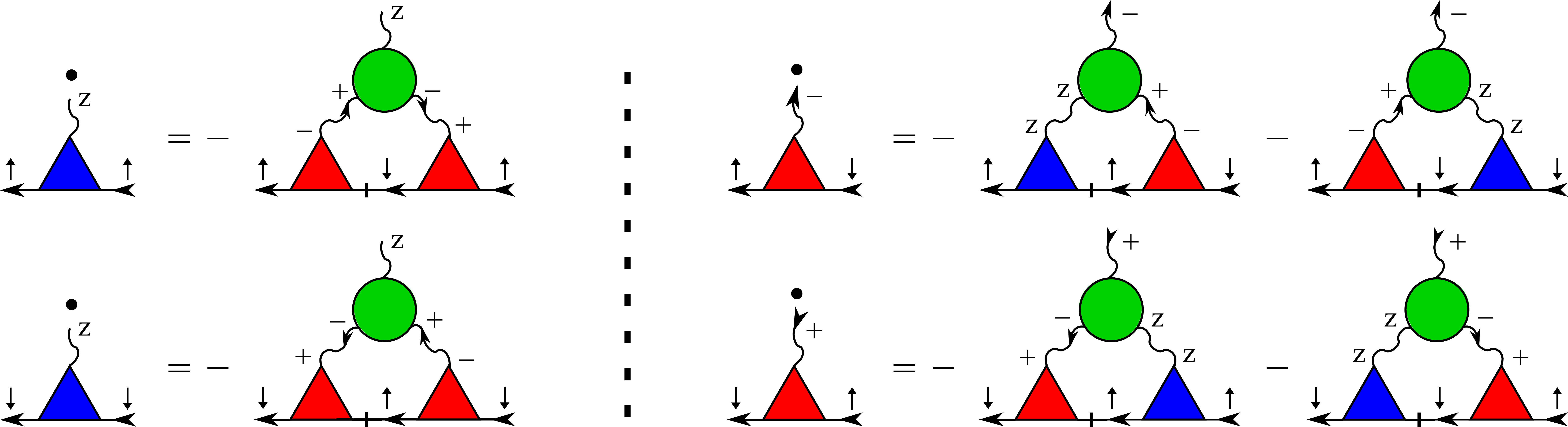}
\caption{
Graphical representation of the leading terms in the 
truncated FRG flow equations for the generalized exchange couplings
(mixed three-legged vertices) shown 
in the first two columns  of Fig.~\ref{fig:vertices}. Here the dot over the diagrams on the left-hand side of the flow equations represents the derivative $\partial_{\Lambda}$ with respect to the cutoff, and the slashed arrows represent fermionic
single-scale propagators.
}
\label{fig:flow}
\end{figure*}

It is now straightforward to write down exact flow equations for the vertices
appearing in the expansion (\ref{eq:Gammaaniso2}). 
These flow equations depend on various higher-order vertices.
Fortunately, in order to
derive Anderson's ``poor man's scaling'' results, it is sufficient to consider only the
flow of the mixed three-legged vertices,
$J^{\parallel}_{\Lambda} ( \bd{k} ,  \omega , \bd{k}^{\prime} ,   
\omega^{\prime} )$ and
$J^{\bot}_{\Lambda} ( \bd{k} ,  \omega , \bd{k}^{\prime} ,   
\omega^{\prime} )$, which
can be identified with the renormalized exchange couplings provided their momentum and frequency dependence can be ignored.
Moreover, in the weak coupling limit we may truncate the flow equations
by retaining only those diagrams which depend quadratically on the
exchange couplings.
The corresponding diagrams are shown graphically in Fig.~\ref{fig:flow}.
Explicitly, the corresponding flow equations
for the generalized exchange couplings are
 \begin{widetext}
\begin{eqnarray}
 \partial_{\Lambda} J^{\parallel}_{\Lambda} ( \bd{k}^{\prime} , \omega^{\prime}, 
 \bd{k} , \omega ) & = & - \frac{1}{N \beta} \sum_{\bd{q}} \sum_{ \nu}    
 J^{\bot}_{\Lambda} ( \bd{k}^{\prime} , \omega^{\prime} , \bd{q} , \omega + \nu )
 \dot{G}_{\Lambda} ( \bd{q} , \omega + \nu )
 J^{\bot}_{\Lambda} ( \bd{q} , \omega + \nu , \bd{k} , \omega )
 \nonumber
 \\
 &  & \hspace{18mm} \times F^{\bot}_{\Lambda} ( \nu ) 
 {\Gamma}^{-+z}_{\Lambda} ( - \nu , \omega - \omega^{\prime} + \nu , \omega^{\prime} - \omega )   F^{\bot}_{\Lambda} ( \omega - \omega^{\prime} + \nu )   ,
 \label{eq:flowJparallel}
 \\
 \partial_{\Lambda} J^{\bot}_{\Lambda} ( \bd{k}^{\prime} , \omega^{\prime}, 
 \bd{k} , \omega ) & = & - \frac{1}{ N \beta} \sum_{\bd{q}} \sum_{ \nu} 
 \bigl[   
 J^{\uparrow \uparrow }_{\Lambda} ( \bd{k}^{\prime} , \omega^{\prime} , \bd{q} , \omega + \nu )
 \dot{G}_{\Lambda} ( \bd{q} , \omega + \nu )
 J^{\bot}_{\Lambda} ( \bd{q} , \omega - \nu , \bd{k} , \omega )
 \nonumber
 \\
 &  & \hspace{23mm} \times F^{\bot}_{\Lambda} ( \nu ) 
 {\Gamma}^{-+z}_{\Lambda} ( \omega^{\prime} - \omega, -\nu,
  \omega - \omega^{\prime} + \nu  )   F^{\parallel}_{\Lambda} 
 ( \omega - \omega^{\prime} + \nu )
 \nonumber
 \\
 &   & \hspace{20mm} +
 J^{\bot}_{\Lambda} ( \bd{k}^{\prime} , \omega^{\prime} , \bd{q} , \omega + \nu )
 \dot{G}_{\Lambda} ( \bd{q} , \omega + \nu )
 J^{\downarrow \downarrow}_{\Lambda} ( \bd{q} , \omega + \nu , \bd{k} , \omega )
 \nonumber
 \\
 &  & \hspace{23mm} \times F^{\parallel}_{\Lambda} ( \nu ) 
 {\Gamma}^{-+z}_{\Lambda} ( \omega^{\prime} - \omega,
  \omega - \omega^{\prime} + \nu , - \nu )   F^{\bot}_{\Lambda} ( \omega - \omega^{\prime} + \nu )  
   \bigr].
 \label{eq:flowJperp}
 \end{eqnarray}
\end{widetext}
Here $\dot{G}_{\Lambda} ( \bd{k}, \omega )$ 
is the electronic single-scale propagator
in the approximation where the electronic self-energy
$\Sigma_{\Lambda} ( \bd{k} , \bd{k}^{\prime} , \omega )$ is neglected,
 \begin{equation}
 \dot{G}_{\Lambda} ( \bd{k} , \omega )  = \frac{
 \partial_{\Lambda} R_{\Lambda} ( \bd{k} )}{ [
 i \omega - \xi_{\bd{k}} - R_{\Lambda} ( \bd{k} )]^2}.
\label{eq:single_scale_propagator}
 \end{equation}
Also neglecting the momentum and frequency dependence of the exchange couplings,
setting the external frequencies $\omega $ and $\omega^{\prime}$ equal 
to zero,\footnote{Although fermionic Matsubara frequencies never vanish for $T > 0$,
	we shall in this work eventually evaluate the frequency sums in the limit $T \rightarrow 0$ so that for our purpose it is sufficient to formally set the
	fermionic frequencies $\omega$ and $\omega^{\prime}$ 
	in Eqs.~(\ref{eq:flowJparallel}) and (\ref{eq:flowJperp}) equal to zero.} and
using $J^{\uparrow \uparrow}_{\Lambda} = J^{\parallel}_{\Lambda}$ 
as well as $J^{\downarrow \downarrow}_{\Lambda} = - J^{\parallel}_{\Lambda}$, 
the above flow equations reduce to
 \begin{eqnarray}
 \partial_{\Lambda} J^{\parallel}_{\Lambda} 
 & = & - \frac{ ( J^{\bot}_{\Lambda} )^2 }{N \beta} 
 \sum_{\bd{q}} \sum_{ \nu}    
 \dot{G}_{\Lambda} ( \bd{q} , \nu ) 
 \nonumber
 \\
 & & \times  [ F^{\bot}_{\Lambda} ( \nu ) ]^2
 {\Gamma}^{-+z}_{\Lambda} ( - \nu ,  \nu , 0 )   ,
 \label{eq:flowJparallelsimp}
 \\
 \partial_{\Lambda} J^{\bot}_{\Lambda} 
 & = &  - \frac{      J^{\bot}_{\Lambda}  J^{\parallel}_{\Lambda}   }{  N \beta} 
 \sum_{\bd{q}} \sum_{ \nu}    
 \dot{G}_{\Lambda} ( \bd{q} , \nu )  
 \nonumber
 \\
 & & \times 2 F^{\bot}_{\Lambda} ( \nu )  F^{\parallel}_{\Lambda} ( \nu ) 
 {\Gamma}^{-+z}_{\Lambda} ( 0 ,  - \nu ,  \nu ).
 \hspace{7mm}
 \label{eq:flowJperpsimp}
 \end{eqnarray}
The three-legged vertex 
${\Gamma}^{-+z}_{\Lambda} ( 0 ,  - \nu ,   \nu )$ appearing in the above
expressions is related to the corresponding
connected spin correlation function $G_{\Lambda}^{+-z} ( \nu , - \nu , 0 )$
via the  usual tree expansion \cite{Kopietz10}, implying
 \begin{align}
   [ F^{\bot}_{\Lambda} ( \nu ) ]^2
 {\Gamma}^{-+z}_{\Lambda} ( - \nu ,  \nu , 0 ) 
  &=   - \frac{G_{\Lambda}^{+-z} ( \nu , - \nu , 0 )}{F^{\parallel}_{\Lambda} (0 )},
 \label{eq:FG1}
 \\
    F^{\bot}_{\Lambda} ( \nu )  F^{\parallel}_{\Lambda} ( \nu ) 
 {\Gamma}^{-+z}_{\Lambda} ( 0 ,  - \nu ,  \nu ) 
&= 
 - \frac{G_{\Lambda}^{+-z} ( 0 , \nu , - \nu  )}{F^{\bot}_{\Lambda} (0 )}.
 \label{eq:FG2}
 \end{align}
In the weak coupling limit, it is sufficient to approximate the inverse spin propagators and the three-point spin correlation functions on the right-hand side of Eqs.~(\ref{eq:FG1}) and (\ref{eq:FG2}) by the corresponding 
expressions describing a free spin with magnitude $S$.
The static spin propagators are then approximated by
 \begin{eqnarray}
 F_0^{\parallel} (0 )  & \approx &   \beta b_0^{\prime} ,
 \\
 F_0^{\bot} (0 )  & \approx &   2 \beta b_0^{\prime} ,
 \end{eqnarray}
where 
 \begin{equation}
 b_0^{\prime} = \frac{ S ( S+1)}{3}
 \end{equation}
can be identified with the derivative of the spin-$S$ Brillouin function at vanishing
magnetic field.
The connected three-point spin correlation function
$ G_{0}^{+-z} ( \nu_1 , \nu_2 , \nu_3  )$ of a single isolated spin
in an external magnetic field has been derived by
Vaks, Larkin, and Pikin \cite{Vaks68}. Here we only need the limit
of vanishing external magnetic field,
 \begin{eqnarray}
  G^{+-z}_0 (  \nu_1 ,  \nu_2 , \nu_3 ) & = &    - 2  \beta b_0^{\prime} 
     ( 1 - \delta_{\nu_1,0} \delta_{\nu_2,0} \delta_{\nu_3, 0 } )  
 \nonumber
 \\
 &    \times &
\left[ \frac{ \delta_{\nu_1,0}  }{ i \nu_2 } +  
\frac{ \delta_{\nu_2 ,0 }  }{ i \nu_3 }
 + \frac{ \delta_{\nu_3 ,0 } }{ i \nu_1 } \right].
 \hspace{7mm}
 \end{eqnarray}
We conclude that
 \begin{align}
  [ F^{\bot}_{0} ( \nu ) ]^2
 {\Gamma}^{-+z}_{0} ( - \nu ,  \nu , 0 )  
 &=
 2 F^{\bot}_{0} ( \nu )  F^{\parallel}_{0} ( \nu ) 
 {\Gamma}^{-+z}_{0} ( 0 ,  - \nu ,  \nu )  
 \nonumber
 \\
 &=   \left\{
 \begin{array}{cc}
 0 & \mbox{for $ \nu =0$,} \\
 2  / ( i \nu ) & \mbox{for $ \nu \neq 0 $,}
 \end{array}
 \right. 
 \label{eq:vertres}
 \end{align}
which is independent of the spin $S$.
Substituting the free-spin result (\ref{eq:vertres}) for the
flowing spin vertices in 
Eqs.~(\ref{eq:flowJparallelsimp}) and (\ref{eq:flowJperpsimp}) we obtain the flow equations
 \begin{eqnarray}
  \partial_{\Lambda} J^{\parallel}_{\Lambda} & = & - A_{\Lambda} 
 ( J_{\Lambda}^{\bot} )^2,
 \\
   \partial_{\Lambda} J^{\bot}_{\Lambda} & = & - A_{\Lambda} 
  J_{\Lambda}^{\bot}  J_{\Lambda}^{\parallel}    ,
 \end{eqnarray} 
where
 \begin{equation}
 A_{\Lambda} = \frac{ 2 }{ \beta N } \sum_{\bd{q}} \sum_{ \nu \neq 0 }
  \frac{\dot{G}_{\Lambda} ( \bd{q} , \nu )}{  i \nu }. 
 \label{eq:Adef}
 \end{equation}
To proceed, we have to specify our regulator.
The bandwidth cutoff scheme adopted by Anderson in his  ``poor man's scaling'' approach
can be implemented in the FRG via the  Litim regulator given in Eq.~(\ref{eq:Litim}),
 which suppresses the
propagation of electrons in an energy shell around the Fermi surface.
 The single-scale propagator as defined in Eq.~\eqref{eq:single_scale_propagator} is then of the form
 \begin{equation}
 \dot{G}_{\Lambda} ( \bd{k} , \omega ) 
= \frac{     ( {\rm sign} \xi_{\bd{k}} ) \Theta (   \Lambda - | \xi_{\bd{k}} | )       }{ [ i \omega - \xi_{\bd{k}} - R_{\Lambda} ( \bd{k} )]^2}.
 \end{equation}
Neglecting the energy dependence of the 
density of states, we can easily carry out the momentum integration
in Eq.~(\ref{eq:Adef}),
 \begin{equation}
 \frac{1}{N} \sum_{\bd{q}} \dot{G}_{\Lambda} ( \bd{q} , \nu ) =  \rho_0 \frac{ 4 i \nu \Lambda^2}{
 ( \nu^2 + \Lambda^2 )^2},
 \end{equation}
where $\rho_0$ is the density of states at the Fermi energy per 
spin projection, see Eq.~(\ref{eq:rho0def}).
In the zero-temperature limit, the Matsubara sum in Eq.~(\ref{eq:Adef}) then becomes elementary and we obtain 
 \begin{equation}
  A_{\Lambda}  = 2 \rho_0 / \Lambda.
 \end{equation}
Introducing the logarithmic flow parameter
 $l = \ln ( \Lambda_0 / \Lambda )$ and using
$\partial_l = -  \Lambda \partial_{\Lambda}$
we finally arrive at the well-known scaling equations 
for the anisotropic Kondo model \cite{Coleman15,Kogan18},
 \begin{eqnarray}
\partial_{l} J^{\parallel}_{\Lambda} & = & 2 \rho_0 
 ( J_{\Lambda}^{\bot} )^2,
 \\
 \partial_{l} J^{\bot}_{\Lambda} & = & 2 \rho_0 
  J_{\Lambda}^{\bot}  J_{\Lambda}^{\parallel}   .
 \end{eqnarray} 
Note that our derivation is valid for arbitrary impurity spin $S$. The fact that the spin cancels
in the weak coupling scaling equations indicates that the Kondo energy
where the system crosses over to the  strong coupling regime is independent of $S$.
This is consistent with the exact Bethe-ansatz solution of the spin-$S$ Kondo 
model \cite{Fateev81}.

\section{Conclusions}

In this work we have shown how to derive the one-loop scaling equations for
the anisotropic Kondo model within the functional renormalization group
approach for quantum spin systems recently proposed in
Ref.~[\onlinecite{Krieg18}]. Of course, the scaling laws for the Kondo model
are well known and
have been derived a long time ago using less sophisticated methods.
Nevertheless, it is conceptually important to show how these scaling laws
can be derived within the framework of the FRG 
because the FRG claims to unify 
different  implementations of the renormalization group idea \cite{Kopietz10,Krieg17}.
In the same way that
the 
 ``poor man's scaling'' approach has superseded the
more complicated 
space-time approach adopted earlier by 
Anderson, Yuval, and Hamann\cite{Anderson70b},
the FRG approach supersedes the 
``poor man's scaling'' approach, since it avoids the
unconventional $T$-matrix renormalization and embeds the  renormalization group theory
for the Kondo model into the established framework of the FRG.
Although it requires some effort to become acquainted with the FRG formalism,
once this is achieved one can 
transcend the ``poor man`s scaling'' approach in several directions. For example, by adopting a cutoff scheme where at the initial scale  
the 
electronic bandwidth vanishes, the spin FRG can also be used to study the
strong coupling regime of the Kondo model, where for $S =1/2$
the impurity spin is completely screened and the conduction electrons interact via an induced two-body 
interaction \cite{Nozieres74}.
Moreover, within the FRG it is straightforward to keep track of the renormalization group flow
of the electronic self-energy $\Sigma_{\Lambda} ( \bd{k} , \bd{k}^{\prime} ,
 \omega )$, which contains information about the spatial distribution of the
 charge density in the vicinity of the impurity spin \cite{Affleck08}.
In principle,  our approach can also be used to 
study multi-channel Kondo models or other impurity models where the strong coupling phase is not a Fermi liquid.
Finally, let us emphasize again that  our spin FRG  approach works directly 
with the physical spin operators, thus avoiding the complications 
which arise if 
the impurity spin is represented in terms of
auxiliary fermions, such as Abrikosov pseudofermions \cite{Abrikosov65,Coleman15,Reuther10} or Majonana fermions \cite{Tsvelik92,Shnirman03}.

\section*{Acknowledgements}
We thank O. Tsyplyatyev for useful discussions. 
One of us (P.K.) acknowledges the hospitality
of the Department of Physics and Astronomy of the University of California, Irvine,
where this work was completed.

\bibliographystyle{apsrev4-1}
\bibliography{kon}

\end{document}